\documentstyle[multicol,pre,aps,epsf,amsmath]{revtex}

\begin{document}

\draft

\title{Fast and stable method for simulating quantum electron dynamics}

\author{Naoki Watanabe, Masaru Tsukada}

\address{Department of Physics,%
  Graduate School of Science, University of Tokyo
  7-3-1 Hongo, 113-0033 Bunkyo-ku, Tokyo, Japan}

\date{Published from Physical Review E. {\bf 62}, 2914, (2000).}

\maketitle

\begin{abstract}
A fast and stable method is formulated to compute the time evolution of
a wavefunction by numerically solving the time-dependent Schr{\"o}dinger
equation. This method is a real space/real time evolution method
implemented by several computational techniques such as Suzuki's exponential
product, Cayley's form, the finite differential method and an operator named
adhesive operator. This method conserves the norm of the wavefunction, manages
periodic conditions and adaptive mesh refinement technique, and is suitable
for vector- and parallel-type supercomputers. Applying this method to some
simple electron dynamics, we confirmed the efficiency and accuracy of the method
for simulating fast time-dependent quantum phenomena.
\end{abstract}

\pacs{02.70.-c,03.67.Lx,73.23,42.65.-k}

\begin{multicols}{2}
\narrowtext

\section{Introduction}

There are many computational method of solving the TD-Schr{\"o}dinger equation
numerically. Conventionally, a wavefunction has been represented as a linear
combination of plane waves or atomic orbitals. However, these representations
entail high computational cost to calculate the matrix elements for these bases.
The plane wave bases set is not suitable for localized orbitals, and the atomic
orbital bases set is not suitable for spreading waves. Moreover, they are not
suitable for parallelization, since the calculation of matrix elements requires
massive data transmission among processors.

To overcome those problems, some numerical methods adopted real space
representation\cite{Varga1962,DeRaedt1994,Iitaka1994,Natori1997}. In those methods,
a wavefunction is descritized by grid points in real space, and with them
some dynamic electron phenomena were simulated successfully
\cite{DeRaedt1994PRB,Iitaka1997,Kono1997}.

Among these real space methods, a method called Cayley's form or
Crank-Nicholson scheme is known to be especially useful for one-dimensional
closed systems because this method conserves the norm of the wavefunction exactly
and the simulation is rather stable and accurate even in a long time slice.
These characteristics are very attractive for simulations over a long time span.
Unfortunately, this method is not suitable for two- or three-dimensional systems.
This problem is fatal for physically meaningful systems. Though there are many
other computational methods that can manage two- or three-dimensional systems,
these methods also have disadvantages.

In the present work, we have overcome the problems associated with Cayley's
form and have formulated a new computational method which is more efficient,
more adaptable and more attractive than any other ordinary methods.

In our method, all computations are performed in real space so there is no
need of using Fourier transform. The time evolution operator in our method is
exactly unitary by using Cayley's form and Suzuki's exponential product
so that the norm of the wavefunction is conserved during the time evolution.
Stability and accuracy are improved by Cayley's form so we can use a longer
time slice than those of the other methods. Cayley's form is a kind of implicit
methods, this is the key to the stability, but implicit methods are not
suitable for periodic conditions and parallelization. We have avoided these
problems by introducing an operator named adhesive operator. This adhesive
operator is also useful for adaptive mesh refinement technique.

Our method inherits many advantages from many ordinary methods, and yet more
improved in many aspects. With these advantages, this method will be useful
for simulating large-scale and long-term quantum electron dynamics
from first principles.

In section II, we formulate the new method step by step. In section III,
we apply it to some simulations of electron dynamics and demonstrate its
efficiency. In section IV, we draw some conclusions.

\section{Formulation}

In this section, we formulate the new method step by step
from the simplest case to complicated cases.
Throughout this paper, we use the atomic units $\hbar=1,\,m=1,\,e=1$.

\subsection{One-dimensional closed free system}

For the first step, we consider a one-dimensional closed system
where an electron moves freely but never leaks out of the system.
The TD-Schr{\"o}dinger equation of this system is simply given as
\begin{equation}
  {\rm i} \frac{\partial \psi(x,t)}{\partial t}
   =
   - \frac{\partial_x^2}{2}\, \psi(x,t)\ .
\label{2-1-1}
\end{equation}

The solution of Eq.~(\ref{2-1-1}) is analytically given by an exponential
operator as
\begin{equation}
  \psi(x,t+\Delta{t})
  =
  \exp{\Bigl[{\rm i}\Delta{t}\frac{\partial_x^2}{2}\Bigr]} \  \psi(x,t)\ ,
\label{2-1-2}
\end{equation}
where $\Delta{t}$ is a small time slice. By using Eq.~(\ref{2-1-2})
repeatedly, the time evolution of the wavefunction is obtained.

An approximation is utilized to make a concrete form of the exponential
operator. We have to be careful not to destroy the unitarity of the time
evolution operator, otherwise the wavefunction rapidly diverges.
We adopted Cayley's form because it is unconditionally stable and accurate
enough. Cayley's form is a fractional approximation of the exponential operator
given by
\begin{equation}
  \exp{\Bigl[{\rm i}\Delta{t}\frac{\partial_x^2}{2}\Bigr]}
  \simeq
  \frac{1+{\rm i}\Delta{t}\partial_x^2/4}{1-{\rm i}\Delta{t}\partial_x^2/4}\ .
\label{2-1-3}
\end{equation}
It is second-order accurate in time.
By substituting Eq.~(\ref{2-1-3}) for Eq.~(\ref{2-1-2}) and
moving the denominator onto the left-hand side, the following
basic equation is obtained:
\begin{equation}
  \Bigl[ 1 - {\rm i} \frac{\Delta{t}}{2} \frac{\partial_x^2}{2} \Bigr]\  \psi(x,t+\Delta{t})
  =
  \Bigl[ 1 + {\rm i} \frac{\Delta{t}}{2} \frac{\partial_x^2}{2} \Bigr]\  \psi(x,t)\ .
\label{2-1-4}
\end{equation}
This is identical with the well-known Crank-Nicholson scheme.
The wavefunction is descritized by grid points in real space as
\begin{equation}
  \psi_i(t) = \psi(x_i,t) \ ;\qquad
  x_i = i\Delta{x},\quad i=0,\cdots,N-1
\end{equation}
where $\Delta{x}$ is the span of the grid points.
We approximate the spatial differential operator by the
finite difference method (FDM).
Then Eq.~(\ref{2-1-4}) becomes a simultaneous linear equation for the vector
quantity $\psi_i(t+\Delta{t})$.
For example, in a system with six grid points, Eq.~(\ref{2-1-4}) is approximated
in the following way:
\begin{multline}
  \left[\begin{array}{cccc}
    A & -1 &  0 &  0 \\
   -1 &  A & -1 &  0 \\
    0 & -1 &  A & -1 \\
    0 &  0 & -1 &  A
  \end{array}\right]
  \left[\begin{array}{c}
    \psi_1(t+\Delta{t}) \\
    \psi_2(t+\Delta{t}) \\
    \psi_3(t+\Delta{t}) \\
    \psi_4(t+\Delta{t}) 
  \end{array}\right] \\
  =
  \left[\begin{array}{cccc}
    B & 1 & 0 & 0 \\
    1 & B & 1 & 0 \\
    0 & 1 & B & 1 \\
    0 & 0 & 1 & B
  \end{array}\right]
  \left[\begin{array}{c}
    \psi_1(t) \\
    \psi_2(t) \\
    \psi_3(t) \\
    \psi_4(t) 
  \end{array}\right]
\label{2-1-5}
\end{multline}
In the above,
\begin{equation}
  A \equiv - 4{\rm i}\frac{\Delta{x}^2}{\Delta{t}} + 2\ , \
  B \equiv - 4{\rm i}\frac{\Delta{x}^2}{\Delta{t}} - 2\
\end{equation}
and $\psi_0$ and $\psi_5$ are fixed at zero due to the boundary condition.

It is easy to solve this simultaneous linear equation because
the matrix appearing on the left-hand side is easily decomposed
into the LU form as
\begin{multline}
  \left[\begin{array}{cccc}
   u_1^{-1} & 0 &  0 &  0 \\
   -1 & u_2^{-1} & 0 &  0 \\
     0 & -1 & u_3^{-1} & 0 \\
     0 &  0 & -1 & u_4^{-1}
  \end{array}\right]
  \left[\begin{array}{cccc}
     1 &-u_1&  0 &  0 \\
     0 &  1 &-u_2&  0 \\
     0 &  0 &  1 &-u_3\\
     0 &  0 &  0 & 1 
  \end{array}\right] \\
\times
  \left[\begin{array}{c}
    \psi_1(t+\Delta{t}) \\
    \psi_2(t+\Delta{t}) \\
    \psi_3(t+\Delta{t}) \\
    \psi_4(t+\Delta{t})
  \end{array}\right]
  =
  \left[\begin{array}{c}
    b_1(t) \\
    b_2(t) \\
    b_3(t) \\
    b_4(t)
  \end{array}\right]
\end{multline}

Here $b_i$ and $u_i$ are auxiliary vectors defined as below
\begin{align}
  b_i(t) &\equiv \psi_{i-1}(t) + B \psi_i(t) + \psi_{i+1}(t), \\
  u_i &\equiv 1/(A-u_{i-1}), \quad u_0 \equiv 0
\label{2-1-6}
\end{align}
The auxiliary vector $u_i$ is determined in advance, and
it is treated as a constant vector in Eq.~(\ref{2-1-6}).
$26N$ floating operations are heeded to solve Eq.~(\ref{2-1-6});
here $N$ is the number of the grid points in the system,
about twice that of the Euler method. Unlike the Euler method,
it exactly conserves the norm because the matrices in Eq.~(\ref{2-1-5})
are unitary. Moreover, the expected energy is conserved because
the time evolution operator commutes with the Hamiltonian in this case.

\subsection{Three-dimensional closed free system}

It is easy to extend this technique to a three-dimensional system.
The formal solution of the TD-Schr{\"o}dinger equation 
in a three-dimensional system is given by
an exponential of the sum of three second differential operators as
\begin{equation}
  \psi({\bf r},t+\Delta{t})
  =
  \exp{\Bigl[{\rm i}\Delta{t} \Bigl(
    \frac{\partial_x^2}{2}
  + \frac{\partial_y^2}{2}
  + \frac{\partial_z^2}{2}
  \Bigr)\Bigr]} \  \psi({\bf r},t)\ .
\label{2-2-1}
\end{equation}
These differential operators in Eq.~(\ref{2-2-1}) are
commutable among each other, so the exponential operator is
exactly decomposed into a product of three exponential operators:
\begin{multline}
  \psi({\bf r},t+\Delta{t})
  =
  \exp{\Bigl[{\rm i}\Delta{t}\frac{\partial_x^2}{2}\Bigr]}\  
  \exp{\Bigl[{\rm i}\Delta{t}\frac{\partial_y^2}{2}\Bigr]}\ \\\times
  \exp{\Bigl[{\rm i}\Delta{t}\frac{\partial_z^2}{2}\Bigr]}\  
  \psi({\bf r},t)\ .
\label{2-2-2}
\end{multline}

Each exponential operator is approximated by Cayley's form as
\begin{multline}
  \psi({\bf r},t+\Delta{t})
  =
  \frac{1+{\rm i}\Delta{t}\partial_x^2/4}{1-{\rm i}\Delta{t}\partial_x^2/4}\cdot
  \frac{1+{\rm i}\Delta{t}\partial_y^2/4}{1-{\rm i}\Delta{t}\partial_y^2/4}\cdot
\\\times
  \frac{1+{\rm i}\Delta{t}\partial_z^2/4}{1-{\rm i}\Delta{t}\partial_z^2/4}\,
  \psi({\bf r},t)\ .
\label{2-2-3}
\end{multline}

$78N$ floating operations are required to compute Eq.~(\ref{2-2-3});
where $N$ is the total number of grid points in the system.
The norm and energy are conserved exactly.

By the way, a conventional method, Peaceman-Rachfold method\cite{Varga1962,Kono1997},
utilizes similar approximation appearing on Eq.~(\ref{2-2-3}),
which is a kind of the alternating direction implicit method (ADI method).
However, by using exponential product,
we have found that there is no need of ADI.
This fact makes the programming code simpler and it runs faster.

\subsection{Static potential}

Next we consider a system subjected to a static external scalar field
$V({\bf r})$.
The TD-Schr{\"o}dinger equation and its formal solution
in this system are as follows:
\begin{gather}
  {\rm i} \frac{\partial \psi({\bf r},t)}{\partial t}
  =
  \Bigl[
    -\frac{\triangle}{2} + V({\bf r})
  \Bigr]\,
  \psi({\bf r},t)\ .
\label{2-3-1}\\
  \psi({\bf r},t+\Delta{t})
  =
  \exp{\Bigl[
    {\rm i}\Delta{t}\frac{\triangle}{2} - {\rm i}\Delta{t} V({\bf r})
  \Bigr]}\ 
  \psi({\bf r},t)\ .
\label{2-3-2}
\end{gather}
To cooperate with the potential in the framework of the formula
described in the previous subsections,
we have to separate the potential operator from the kinetic operator
using Suzuki's exponential product theory\cite{Suzuki1990,Suzuki1991} as
\begin{multline}
  \psi({\bf r},t+\Delta{t})
  =
  \exp{\Bigl[-{\rm i}\frac{\Delta{t}}{2}V\Bigr]}\ 
  \exp{\Bigl[ {\rm i}\Delta{t}\frac{\triangle}{2}\Bigr]}\  \\\times
  \exp{\Bigl[-{\rm i}\frac{\Delta{t}}{2}V\Bigr]}\
  \psi({\bf r},t)\ .
\label{2-3-3}
\end{multline}

\begin{figure}[htbp]
  \begin{center}
    \epsfxsize=70mm\mbox{\epsfbox{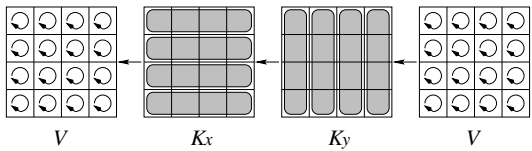}}
  \end{center}
  \caption{%
    The procedure for a two-dimensional closed system with a static potential.
    Here $V$ shows the operation of the exponential of the potential,
    which changes the phase of the wavefunction at each grid point.
    $K_x$ and $K_y$ show the operation of Cayley's form along the x-axis
    and the y-axis respectively. They are computed independently
    by grid points, by rows, or by columns.
  }
  \label{fig2-3-1}
\end{figure}

This decomposition is correct up to the second-order of $\Delta{t}$.
The exponential of the potential is computed by just changing
the phase of the wavefunction at each grid point.
The exponential of the Laplacian is computed in the way 
described in the previous subsections.
Each operator is exactly unitary, so the norm is
conserved exactly. But due to the separation of the incommutable
operators, the energy is not conserved exactly.
Yet it oscillates near around its initial values and it
never drifts monotonously.
This algorithm is quite suitable for vector-type supercomputers
because all operations are independent by grid points, by rows, or by columns.
The outline of this procedure for a two-dimensional system
is schematically described by Fig.~\ref{fig2-3-1}.

The decomposition (\ref{2-3-3}) is a second-order one.
Higher-order decompositions are derived using Suzuki's fractal
decomposition\cite{Suzuki1990,Suzuki1991,Umeno1993,Suzuki1993springer}.
For instance, a fourth-order fractal decomposition ${\bf S}_{4}(\Delta{t})$ is
given by
\begin{multline}
  {\bf S}_{4}(\Delta{t})
  =
  {\bf S}_2(s\Delta{t})\ 
  {\bf S}_2(s\Delta{t})\
  {\bf S}_2((1-4s)\Delta{t})\ \\\times
  {\bf S}_2(s\Delta{t})\
  {\bf S}_2(s\Delta{t})
\label{2-3-4}
\end{multline}
where
\begin{gather}
  {\bf S}_2(\Delta{t})
  \equiv
  \exp{\Bigl[-{\rm i}\frac{\Delta{t}}{2}V\Bigr]}
  \exp{\Bigl[ {\rm i}\Delta{t}\frac{\triangle}{2}\Bigr]}
  \exp{\Bigl[-{\rm i}\frac{\Delta{t}}{2}V\Bigr]}\,\notag\\[3mm]
  s \equiv 1/(4-\sqrt[3]{4})\ .
\label{2-3-5}
\end{gather}

\subsection{Dynamic potential}

To discuss high-speed electron dynamics caused by a time-dependent
external field $V({\bf r},t)$, 
we should take account of the evolution of the potential itself in
the TD-Schr{\"o}dinger equation given as
\begin{equation}
  {\rm i} \frac{\partial \psi({\bf r},t)}{\partial t}
  =
  {\cal H}(t)\, \psi({\bf r},t)\ ;
\quad
  {\cal H}(t)
  =
  -\frac{\triangle}{2} + V({\bf r},t)\ .
\label{2-4-1}
\end{equation}
The analytic solution of Eq.~(\ref{2-4-1}) is given by a Dyson's
time ordering operator ${\bf P}$ as
\begin{equation}
  \psi({\bf r},t+\Delta{t})
  =
  {\bf P} \exp{\Biggl[
    {\rm i}\int_{t}^{t+\Delta{t}}\!\!\!\!\!\!\!\!\!\!\! {\rm d}t^\prime\,\Bigl\{
    \frac{\triangle}{2} - V({\bf r},t^\prime) \Bigr\}
  \Biggr]}
  \psi({\bf r},t)\ .
\label{2-4-8}
\end{equation}

The theory of the decomposition of an exponential with time ordering
was derived by Suzuki\cite{Suzuki1993}. The result is rather simple.
For instance, the second-order decomposition is simply given by
\begin{multline}
  \psi({\bf r},t+\Delta{t})
  \simeq
  \exp{\Bigl[
    - {\rm i} \frac{\Delta{t}}{2} V({\bf r},t+\frac{\Delta{t}}{2})
  \Bigr]}
  \exp{\Bigl[   {\rm i} \Delta{t}\frac{\triangle}{2} \Bigr]} \\\times
  \exp{\Bigl[
    - {\rm i} \frac{\Delta{t}}{2} V({\bf r},t+\frac{\Delta{t}}{2})
  \Bigr]}\
  \psi({\bf r},t)
\label{2-4-13}
\end{multline}
and the fourth-order fractal decomposition is given by
\begin{align}
  \psi({\bf r},t+\Delta{t})
  &=      {\bf S}_2(s\Delta{t};t+(1-s)\Delta{t})   \notag\\
  &\times {\bf S}_2(s\Delta{t};t+(1-2s)\Delta{t})  \notag\\
  &\times {\bf S}_2((1-4s)\Delta{t};t+2s\Delta{t}) \notag\\
  &\times {\bf S}_2(s\Delta{t};t+s\Delta{t})       \notag\\
  &\times {\bf S}_2(s\Delta{t};t) \ \psi({\bf r},t)\ ,
\label{2-4-14}
\end{align}
\begin{multline}
  {\bf S}_2(\Delta{t};t)
  \equiv
  \exp{\Bigl[
    -{\rm i}\frac{\Delta{t}}{2} V({\bf r},t+\frac{\Delta{t}}{2})
  \Bigr]}
  \exp{\Bigl[ {\rm i}\Delta{t}\frac{\triangle}{2}\Bigr]} \\\times
  \exp{\Bigl[
    -{\rm i}\frac{\Delta{t}}{2} V({\bf r},t+\frac{\Delta{t}}{2})
  \Bigr]}\ .
\label{2-4-15}
\end{multline}

These operators are also unitary.
These procedures are quite similar to those of the static potential
except that we take the dynamic potential at the specified time.

\subsection{Periodic system}

In a crystal or periodic system,
the wavefunctions must obey a periodic condition:
\begin{equation}
  \psi({\bf r}+{\bf R},t)
=
  \psi({\bf r},t)\exp{[{\rm i}\phi]}\ ,\qquad \phi \equiv {\bf k}\cdot{\bf R}\ ,
\label{2-6-1}
\end{equation}
where ${\bf k}$ is the Bloch wave number and ${\bf R}$ is the unit vector
of the lattice. The matrix form equation corresponding to Eq.~(\ref{2-1-5})
in this system takes the following form:
\begin{multline}
  \left[\begin{array}{cccc}
    A & -1 &  0 &  e^{+{\rm i}\phi} \\
   -1 &  A & -1 &  0 \\
    0 & -1 &  A & -1 \\
    e^{-{\rm i}\phi} &  0 & -1 &  A
  \end{array}\right]
  \left[\begin{array}{c}
    \psi_1(t+\Delta{t}) \\
    \psi_2(t+\Delta{t}) \\
    \psi_3(t+\Delta{t}) \\
    \psi_4(t+\Delta{t}) 
  \end{array}\right]  \\
  =
  \left[\begin{array}{cccc}
    B & 1 & 0 & e^{-{\rm i}\phi} \\
    1 & B & 1 & 0 \\
    0 & 1 & B & 1 \\
    e^{+{\rm i}\phi} & 0 & 1 & B
  \end{array}\right]
  \left[\begin{array}{c}
    \psi_1(t) \\
    \psi_2(t) \\
    \psi_3(t) \\
    \psi_4(t) 
  \end{array}\right]
\label{2-6-2}
\end{multline}
These matrices have extra elements,
so the equation can no longer be solve efficiently.

We propose a trick to avoid this problem. We represent the second spatial
differential operator $\partial^2_{x}$ as a sum of two operators:
\begin{equation}
  \partial^2_{x} = \partial^2_{x\rm td} + \partial^2_{x\rm ad}\ .
\label{2-6-3}
\end{equation}
Multiplying by $\Delta{x}^2$, the above representation
reads in the matrix form:
\begin{multline}
  \left[\begin{array}{cccc}
    -2 & 1 & 0 & e^{-{\rm i}\phi} \\
    1 & -2 & 1 & 0 \\
    0 & 1 & -2 & 1 \\
    e^{+{\rm i}\phi} & 0 & 1 & -2
  \end{array}\right] \\
=
  \left[\begin{array}{cccc}
    -1 & 1 & 0 & 0 \\
    1 & -2 & 1 & 0 \\
    0 & 1 & -2 & 1 \\
    0 & 0 & 1 & -1
  \end{array}\right]
+
  \left[\begin{array}{cccc}
    -1 & 0 & 0 & e^{-{\rm i}\phi} \\
    0 & 0 & 0 & 0 \\
    0 & 0 & 0 & 0 \\
    e^{+{\rm i}\phi} & 0 & 0 & -1
  \end{array}\right]\ .
\label{2-6-4}
\end{multline}
The first matrix on the right-hand side, which corresponds
to $\partial^2_{x\rm td}$, is tri-diagonal,
and the second one, which corresponds to $\partial^2_{x\rm ad}$,
is its remainder,  and it has a quite simple form.
The exponential of the second differential operator
is decomposed by these terms:
\begin{equation}
  \exp\!\!{\Bigl[\frac{{\rm i}\Delta{t}}{2} \partial^2_{x} \!\Bigr]}
  =
  \exp\!\!{\Bigl[\frac{{\rm i}\Delta{t}}{4} \partial^2_{x\rm ad} \!\Bigr]}
  \exp\!\!{\Bigl[\frac{{\rm i}\Delta{t}}{2} \partial^2_{x\rm td} \!\Bigr]}
  \exp\!\!{\Bigl[\frac{{\rm i}\Delta{t}}{4} \partial^2_{x\rm ad} \!\Bigr]}\ .
\label{2-6-5}
\end{equation}
The exponential of $\partial^2_{x\rm ad}$ is exactly calculated by
the following formula:
\begin{equation}
  \exp{\Bigl[{\rm i}C
  \Bigl(\begin{array}{cc}
    -1 & e^{-{\rm i}\phi} \\
    e^{+{\rm i}\phi} & -1
  \end{array}\Bigr)
\Bigr]}
  =
  {\bf I} + \frac{ 1-e^{-2{\rm i}C} }{2}
  \Bigl(\begin{array}{cc}
    -1 & e^{-{\rm i}\phi} \\
    e^{+{\rm i}\phi} & -1
  \end{array}\Bigr)\ .
\label{2-6-6}
\end{equation}
This operation is exactly unitary and easy to compute.

\begin{figure}[htbp]
  \begin{center}
    \epsfxsize=80mm\mbox{\epsfbox{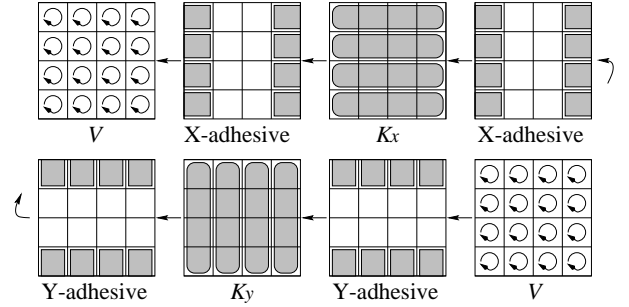}}
  \end{center}
  \caption{
    The procedure for a two-dimensional periodic system.
    Here $K_x$ and $K_y$ show the operations of Cayley's form,
    and they operate as if this system is not periodic.
    X-adhesive and Y-adhesive mean the operations of the exponential of
    the adhesive operators along the x-axis and the y-axis, respectively.
    The operation of the adhesive operator needs only the values at
    the edges of the system.
  }
  \label{fig2-6-1}
\end{figure}

The exponential of $\partial^2_{x\rm td}$ is computed
in the ordinary way. Thus the norm is conserved.
We named $\partial^2_{\rm ad}$ an ``adhesive operator'' because this
operator plays the role of an adhesion to connect both edges of the system.
The outline of the procedure for a two-dimensional periodic
system is schematically described by Fig.~\ref{fig2-6-1}.

\subsection{Parallelization}

\begin{figure}[htbp]
  \begin{center}
    \epsfxsize=80mm\mbox{\epsfbox{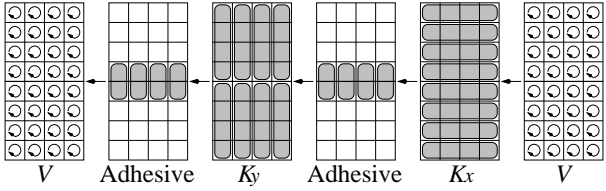}}
  \end{center}
  \caption{
    The procedure for a two-dimensional closed system on two processors.
    Adhesive shows the operation of the exponential of
    the adhesive operator for parallel computing.
    The operation of the adhesive operator needs only the values at
    the edges of the areas, so the data transmission between the processors
    is quite small.
   }
  \label{fig2-7-1}
\end{figure}

The adhesive operator plays another important role. It makes Cayley's
form suitable for parallelization. We use the adhesive operator to
represent the second finite difference matrix in the following way:
\begin{multline}
  \left[\begin{array}{cccc}
    -2 & 1 & 0 & 0 \\
    1 & -2 & 1 & 0 \\
    0 & 1 & -2 & 1 \\
    0 & 0 & 1 & -2
  \end{array}\right] \\
=
  \left[\begin{array}{cccc}
    -2 & 1 & 0 & 0 \\
    1 & -1 & 0 & 0 \\
    0 & 0 & -1 & 1 \\
    0 & 0 & 1 & -2
  \end{array}\right]
+
  \left[\begin{array}{cccc}
    0 & 0 & 0 & 0 \\
    0 & -1 & 1 & 0 \\
    0 & 1 & -1 & 0 \\
    0 & 0 & 0 & 0
  \end{array}\right]\ .
\label{2-7-1}
\end{multline}
The interior of the first matrix on the right-hand side is separated into
two blocks, which means this system is separated into two physically independent
areas. The second matrix, which is the adhesive operator, connects
the two areas.
A large system is separated into many small areas, and each area
is managed by a single processor.
Since the exponential of a block diagonal matrix is also
a block diagonal matrix, each block is computed by a single processor
independently.
Data transmission is needed only to compute the adhesive operator.
The amount of data transmission is quite small, nearly negligible.
The outline of the procedure for a two-dimensional closed
system on two processors
is schematically described by Fig.~\ref{fig2-7-1}.

\subsection{Adaptive mesh refinement}

\begin{figure}[htbp]
  \begin{center}
    \epsfxsize60mm\mbox{\epsfbox{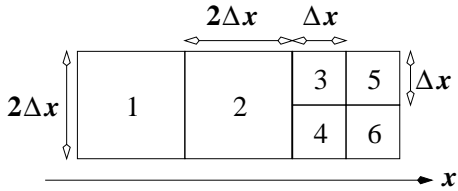}}
  \end{center}
  \caption{
    An example of adaptive mesh refinement.
    The element in the left area is twice as large as
    that in the right area.
    The adhesive operator connects these areas.
  }
  \label{fig2-8-1}
\end{figure}

It is necessary for real space computation to be equipped with
an adaptive mesh refinement to reduce the computational cost or
to improve the accuracy in some important regions.
We improved the adhesive operator
to manage a connection of between two regions whose mesh sizes are
different, as illustrated in Fig.~\ref{fig2-8-1}.

The second differential operator $\partial_{x}^2$
should be Hermite, but in this case the condition required for 
the matrix representation $(\partial_{x}^2)_{ij}$ is given by
\begin{equation}
  (\partial_{x}^2)_{ij} \Delta{x}_i^2
  =
  (\partial_{x}^2)_{ji} \Delta{x}_j^2\ ;
  \qquad
  \text{for all $i, j$}.
\end{equation}

Considering this condition, an approximation of the second differential operator 
is given as
\begin{equation}
  \partial_{x\rm \hskip2ex}^2 = \frac{1}{\Delta{x}^2}
  \begin{tabular*}{56mm}{@{\extracolsep{\fill}}|p{8mm}|p{8mm}|p{8mm}|p{8mm}|p{8mm}|p{8mm}|p{8mm}} \cline{1-6}
   -1/2 &  1/4 &      &      &      &      & {1} \\ \cline{1-6}
    1/4 & -1/2 &  1/8 &  1/8 &      &      & {2} \\ \cline{1-6}
        &  1/2 & -3/2 &      &    1 &      & {3} \\ \cline{1-6}
        &  1/2 &      & -3/2 &      &    1 & {4} \\ \cline{1-6}
        &      &    1 &      &   -2 &      & {5} \\ \cline{1-6}
        &      &      &    1 &      &   -2 & {6} \\ \cline{1-6}
  \end{tabular*}
\label{2-8-1}
\end{equation}
The indices attached to this matrix indicate the corresponding mesh indices
described in Fig.~\ref{fig2-8-1}.
This matrix is also divided into a block-diagonal one and
an adhesive operator as
\begin{equation}
  \partial_{x\rm bd}^2 = \frac{1}{\Delta{x}^2}
  \begin{tabular*}{56mm}{@{\extracolsep{\fill}}|p{8mm}|p{8mm}|p{8mm}|p{8mm}|p{8mm}|p{8mm}|p{8mm}} \cline{1-6}
   -1/2 &  1/4 &      &      &      &      & {1} \\ \cline{1-6}
    1/4 & -1/4 &      &      &      &      & {2} \\ \cline{1-6}
        &      &   -1 &      &    1 &      & {3} \\ \cline{1-6}
        &      &      &   -1 &      &    1 & {4} \\ \cline{1-6}
        &      &    1 &      &   -2 &      & {5} \\ \cline{1-6}
        &      &      &    1 &      & -2   & {6} \\ \cline{1-6}
  \end{tabular*}
\label{2-8-2}
\end{equation}
\begin{equation}
  \partial_{x\rm ad}^2 = \frac{1}{\Delta{x}^2}
  \begin{tabular*}{56mm}{@{\extracolsep{\fill}}|p{8mm}|p{8mm}|p{8mm}|p{8mm}|p{8mm}|p{8mm}|p{8mm}} \cline{1-6}
        &      &      &      &      &      & {1} \\ \cline{1-6}
        & -1/4 &  1/8 &  1/8 &      &      & {2} \\ \cline{1-6}
        &  1/2 & -1/2 &      &      &      & {3} \\ \cline{1-6}
        &  1/2 &      & -1/2 &      &      & {4} \\ \cline{1-6}
        &      &      &      &      &      & {5} \\ \cline{1-6}
        &      &      &      &      &      & {6} \\ \cline{1-6}
  \end{tabular*}
\label{2-8-3}
\end{equation}

The exponential of the adhesive operator is calculated
using the following formula:
\begin{multline}
  \exp{\left[\frac{{\rm i}\Delta{t}}{4\Delta{x}^2}
  \left(\begin{array}{ccc}
   -1/4&  1/8&  1/8\\
    1/2& -1/2&    0 \\
    1/2&    0& -1/2\\
  \end{array}\right)
  \right]}
= \\
  {\bf I}
+
  \left(\begin{array}{ccc}
     2c_1 & -c_1     & -c_1     \\
    -4c_1 & 2c_1+c_2 & 2c_1-c_2 \\
    -4c_1 & 2c_1-c_2 & 2c_1+c_2
  \end{array}\right)\ ,
\label{2-8-4}
\end{multline}
where
\begin{align}
  c_1 &\equiv \frac{1}{6} \exp{\Bigl[
          - \frac{3{\rm i}}{\sqrt{2}} \frac{\Delta{t}}{8\Delta{x}^2}
        \Bigr]} - \frac{1}{6}\ , \\
  c_2 &\equiv \frac{1}{6} \exp{\Bigl[
          - \frac{2{\rm i}}{\sqrt{2}} \frac{\Delta{t}}{8\Delta{x}^2}
        \Bigr]} - \frac{1}{6}\ .
\label{2-8-5}
\end{align}

In this way, it is found that the adhesive operator
is important to simulate a larger or a more complicated system
by the present method.

\section{Application}

In this section, we show some applications of our numerical method.
Though these applications treat simple physical systems, they are
sufficient for verifying the reliability and efficiency of the method.
Throughout this section, we use the atomic units (a.u.).

\subsection{Comparison with conventional methods}

As far as we know, the conventional methods of solving the TD-Schr{\"o}dinger
equation are classified into three categories:
1) the multistep method\cite{Iitaka1994},
2) the method developed by De~Raedt\cite{DeRaedt1994}
and 3) the method equipped with Cayley's form\cite{Recipes}.

\begin{figure}[htbp]
  \begin{center}
    \epsfxsize80mm\mbox{\epsfbox{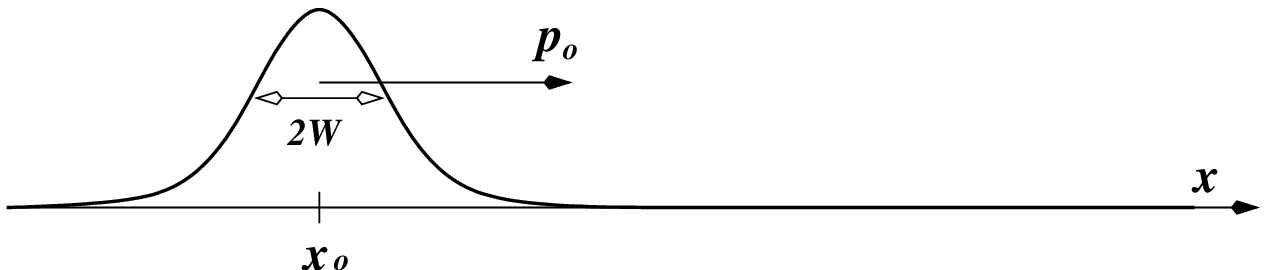}}%
  \end{center}
  \caption{
    The model system for comparison with conventional methods.
    $256$ computational grid points are allocated in the physical length
    $8.0 \text{a.u.}$
    A Gaussian wave packet is placed in the system, whose initial
    average location $x_o$ and momentum $p_o$ are set at
    $x_o=2.0\text{a.u.}$ and $p_o=12.0\text{a.u.}$, respectively.
  }
  \label{fig3-1-0}
\end{figure}

In this section, we make brief comparisons between Cayley's form
and other conventional methods
by simply simulating a Gaussian wave packet moving in a one-dimensional
free system as illustrated in Fig.~\ref{fig3-1-0}.

The TD-Schr{\"o}dinger equation of this system is simply given by
\begin{equation}
  {\rm i}\frac{\partial \psi(x,t)}{\partial t}
=
  - \frac{\partial_x^2}{2}\ \psi(x,t)\ .
  \label{3-1-0}
\end{equation}

The wavefunction at the initial state is set as a Gaussian:
\begin{equation}
  \psi(x,t=0)
=
  \frac{1}{\sqrt[4]{2\pi W^2}}
  \exp{\Bigl[
     -\frac{|x-x_o|^2}{4W^2} + {\rm i} p_o x
  \Bigr]}\ ,
  \label{3-1-1}
\end{equation}
where $W=0.25\text{a.u.},\ x_o=2.0\text{a.u.},\ p_o=12.0\text{a.u.}$

The evolution of this Gaussian is analytically derived as
\begin{multline}
  \psi(x,t)
=
  \frac{1}{\sqrt[4]{2\pi W^2 + (\pi/2)(t/W)^2}} \\\times
  \exp{\Bigl[
    - \frac{(x-x_o-p_o t)^2}{4W^2+(t/W)^2}
    + {\rm i} p_o x
  \Bigr]}\ .
  \label{3-1-2}
\end{multline}
Therefore, the average location of the Gaussian $\langle x(t) \rangle$
is derived as if it is a classical particle:
\begin{equation}
  \langle x(t) \rangle 
=
  \langle x(t=0) \rangle + p_o t\ .
  \label{3-1-3}
\end{equation}
This characteristic is useful to check the accuracy of the simulation.

We use the second-order version of the multistep method and the De~Raedt's method
in order to compare with Cayley's form
since Cayley's form is second-order accurate in space and time.

The second-order multistep method we used in this system is given by
\begin{equation}
  \psi(t+\Delta{t})
=
  \psi(t-\Delta{t}) + {\rm i}2\Delta{t} \frac{\partial_x^2}{2}\, \psi(t)\ ,
  \label{3-1-4}
\end{equation}
where $\partial_x^2$ is approximated by a finite difference matrix as
\begin{equation}
  \partial_{x}^2
  \simeq
  \frac{1}{\Delta{x}^2}
  \left[\begin{array}{cccccc}
   -2 & 1 & 0 & 0 & 0 & 0 \\
    1 &-2 & 1 & 0 & 0 & 0 \\
    0 & 1 &-2 & 1 & 0 & 0 \\
    0 & 0 & 1 &-2 & 1 & 0 \\
    0 & 0 & 0 & 1 &-2 & 1 \\
    0 & 0 & 0 & 0 & 1 &-2
  \end{array}\right]\ .
  \label{3-1-5}
\end{equation}

Extra memories are needed for the wavefunction
at the previous time step $\psi(t-\Delta{t})$.
Though the time evolution of this method is not unitary,
the norm of the wavefunction is conserved with good accuracy
on the condition that $\Delta{t}/\Delta{x}^2 \le 0.5$.
This method needs only $10N$ floating operations per time step,
which is the fastest method in conditionally stable methods.

Meanwhile, the second-order De~Raedt's method is given by
\begin{equation}
  \psi(t+\Delta{t})
=
  \exp\!\!{\Bigl[\frac{{\rm i}\Delta{t}}{2}\frac{\partial_{xa}^2}{2}\!\Bigr]}\!
  \exp\!\!{\Bigl[      {\rm i}\Delta{t}    \frac{\partial_{xb}^2}{2}\!\Bigr]}\!
  \exp\!\!{\Bigl[\frac{{\rm i}\Delta{t}}{2}\frac{\partial_{xa}^2}{2}\!\Bigr]}
  \psi(t)
  \label{3-1-6}
\end{equation}
where $\partial_{xa}^2$ and $\partial_{xb}^2$ are
the parts of the second differential operator
and are approximated by finite difference matrices as below:
\begin{align}
  \partial_{xa}^2
  &\simeq
  \frac{1}{\Delta{x}^2}
  \left[\begin{array}{cccccc}
   -1& 1 & 0 & 0 & 0 & 0 \\
    1 &-1& 0 & 0 & 0 & 0 \\
    0 & 0 &-1& 1 & 0 & 0 \\
    0 & 0 & 1 &-1& 0 & 0 \\
    0 & 0 & 0 & 0 &-1& 1 \\
    0 & 0 & 0 & 0 & 1 &-1
  \end{array}\right]\ ,\\
  \partial_{xb}^2
  &\simeq
  \frac{1}{\Delta{x}^2}
  \left[\begin{array}{cccccc}
   -1& 0 & 0 & 0 & 0 & 0 \\
    0 &-1& 1 & 0 & 0 & 0 \\
    0 & 1 &-1& 0 & 0 & 0 \\
    0 & 0 & 0 &-1& 1 & 0 \\
    0 & 0 & 0 & 1 &-1& 0 \\
    0 & 0 & 0 & 0 & 0 &-1
  \end{array}\right]\ .
  \label{3-1-7}
\end{align}

The exponentials of those matrices are exactly calculated
using the following formula:
\begin{equation}
  \exp{\Bigl[{\rm i}C \begin{pmatrix} -1 & 1 \\ 1 & -1 \end{pmatrix} \Bigr]}
=
  {\bf I} + \frac{ 1- e^{-2{\rm i}C} }{2}
  \begin{pmatrix} -1 & 1 \\ 1 & -1 \end{pmatrix}\ .
  \label{3-1-8}
\end{equation}
The time evolution of this method is exactly unitary,
and the norm is exactly conserved unconditionally.
However, it seems that the accuracy tends to break down 
on the condition that $\Delta{t}/\Delta{x}^2 > 1.0$.
This method needs $18N$ floating operations per time step,
which is the fastest method in unconditionally norm-conserving methods.

Cayley's form with the finite difference method is given by
\begin{equation}
  \psi(t+\Delta{t})
=
  \frac{1 + {\rm i}\Delta{t}/4\ \partial_{x}^2}
       {1 - {\rm i}\Delta{t}/4\ \partial_{x}^2}\,
  \psi(t)\ ,
  \label{3-1-9}
\end{equation}
where the spatial differential operator is approximated by the ordinary way
in Eq.~(\ref{3-1-5}).

The time evolution of this method is exactly unitary, 
and the norm is exactly conserved unconditionally.
Moreover, this method maintains good accuracy even under the condition
that $\Delta{t}/\Delta{x}^2 > 1.0$.
This method needs $26N$ floating operations per time step,
which is the fastest method in unconditionally stable methods.

\begin{figure}[htbp]
  \begin{center}
    \epsfxsize80mm\mbox{\epsfbox{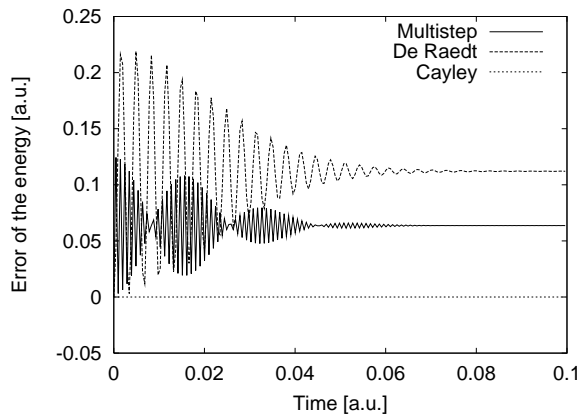}}
  \end{center}
  \caption{
    Time variances in the energies computed by the three methods.
    The time slice is set at $\Delta{t}=1/2048\text{a.u.}$
    and the spatial slice is set at $\Delta{x}=1/32\text{a.u.}$
    so that the ratio $\Delta{t}/\Delta{x}^2$ is equal to $0.5$.
    The energies violently oscillates in the result of
    the multistep method and De~Raedt's method. Meanwhile,
    the energy is conserved exactly in the result of Cayley's form.
  }
  \label{energy}
\end{figure}

We have simulated the motion of the Gaussian by those methods.
First we show a comparison of Cayley's form with the conventional
methods in the framework of the FDM.
Figure~\ref{energy} shows the time evolution of the error in the energy,
which is evaluated by the finite difference method as described below
\begin{gather}
  \epsilon(t) = E(t) - E(t=0)
  \label{3-1-12}\\
  E(t)
  =
  -\frac{1}{2\Delta{x}} {\rm Re} \sum_{i=0}^{N-1}
  \psi_i^{\ast}(t)
  \bigl( \psi_{i-1}(t) - 2\psi_i(t) + \psi_{i+1}(t) \bigr)\ .
\end{gather}
The initial energy is evaluated as $73.03\text{a.u.}$,
though it is theoretically expected to be $74\text{a.u.}$
The ratio $\Delta{t}/\Delta{x}^2$ is set at $0.5$ to
meet the stable condition required for the multistep method.

The energies violently oscillate in the results of
the multistep method and De~Raedt's method, as a result of the fact
that these time evolution operators do not commute with the Hamiltonian. 
These energies seem to converge after the wave packet is delocalized in a
uniform way over the system.
Meanwhile, the energy is conserved exactly in the result of Cayley's form
because Cayley's form commutes with the spatial second differential
operator which is the Hamiltonian itself in this system.

\begin{figure}[htbp]
  \begin{center}
    \epsfxsize80mm\mbox{\epsfbox{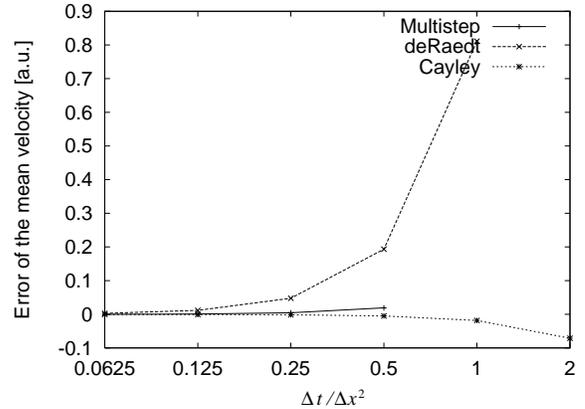}}
  \end{center}
  \caption{
    Errors in the average momentum computed by the three methods in
    several time slices. The multistep method cannot be
    performed when $\Delta{t}/\Delta{x}^2>0.5$. The error of De~Raedt's
    method is too large when $\Delta{t}/\Delta{x}^2>1$.
    The error of Cayley's form is rather small.
    The spatial slice is set at $\Delta{x}=1/32\text{a.u.}$
  }
  \label{accuracy}
\end{figure}

Figure~\ref{accuracy} shows the relation of
the time slice $\Delta{t}$ to the error in the average momentum of the
Gaussian,
which is evaluated by the finite difference method as described below:
\begin{gather}
  \epsilon(\Delta{t}/\Delta{x}^2)
  =
  \frac{
    \langle x(t=T) \rangle - \langle x(t=0) \rangle
  }{T} - \langle p(t=0) \rangle\ ,
  \label{3-1-13} \\
  \langle x(t) \rangle
  =
  \Delta{x} \sum_{i=0}^{N-1}
  x_i |\psi_i(t)|^2 \\
  \langle p(t) \rangle
  =
  \frac{1}{2} {\rm Im}
  \sum_{i=0}^{N-1}
  \psi_i(t)^{\ast} \bigl( \psi_{i+1}(t) - \psi_{i-1}(t) \bigr)\ ,
\end{gather}
where $T$ is a time span set at $0.4\text{a.u.}$
The initial momentum $\langle p(t=0) \rangle$ is calculated
as $11.7\text{a.u.}$, which is different from the theoretical value
$p_o=12.0\text{a.u.}$ due to the finite difference method.

In the multistep method,
the computation cannot be performed due to a floating exception,
if the ratio $\Delta{t}/\Delta{x}^2$ exceeds $0.5$.
In De~Raedt's method,
the error becomes too large to plot in this graph if the ratio
$\Delta{t}/\Delta{x}^2$ exceeds $1.0$.
Meanwhile, in Cayley's form, 
the error is not so large even if the ratio
$\Delta{t}/\Delta{x}^2$ exceeds $1.0$.

In this way, Cayley's form is found rather stable.
Therefore, we can use a longer time slice than those of the other methods.
And this Cayley's form becomes suitable for three-dimensional systems,
potentials, periodic conditions, adaptive mesh refinement, and parallelizations
by our improvements in this paper.

\subsection{Test of the adhesive operator}

\begin{figure}[htbp]
  \begin{center}
    \epsfxsize=60mm\mbox{\epsfbox{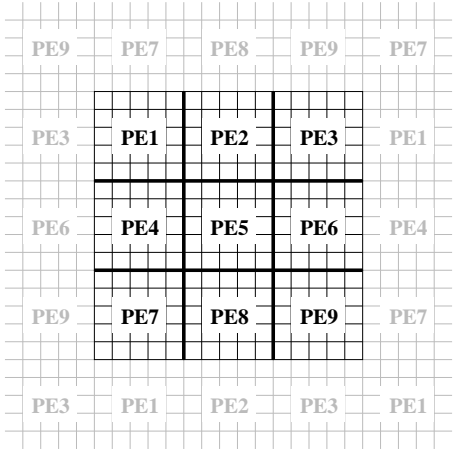}}%
  \end{center}
  \caption{
    The model system for the test of the adhesive operator
    for periodic conditions and parallelization.
    This system is periodically connected and is divided into nine areas.
    Each area is managed by a single processing element.
    $32\times 32$ computational grid points are allocated in each area
    whose physical size is set at $8.0\text{a.u.}\times 8.0\text{a.u.}$
    The time slice is set at $\Delta{t}=1/16\text{a.u.}$
  }
  \label{fig3-2-1}
\end{figure}

To verify the reliability and efficiency
of the adhesive operator for periodic
condition and parallelization,
we have simulated the motion of a Gaussian wave packet in
a two-dimensional free system. As illustrated in Fig.~\ref{fig3-2-1},
this system has periodic conditions along both the x-axis and the y-axis,
and it is divided into nine areas, each of them is managed by a single processing
element; the adhesive operator connects them.
The initial wavefunction is set as a Gaussian given as
\begin{equation}
  \psi({\bf r},t=0)
  =
  \frac{1}{\sqrt{2\pi W^2}}
  \exp{\Bigl[ -\frac{|{\bf r}-{\bf r_o}|^2}{4W^2}
    + {\rm i}{\bf p_o}\cdot{\bf r} \Bigr]}\ ,
  \label{3-2-1}
\end{equation}
where ${\bf r_o}$ is set as the center of this system
and ${\bf p_o}=(1\text{a.u.},1\text{a.u.})$, $W=1\text{a.u.}$
The energy of this Gaussian is theoretically derived as $1.0625\text{a.u.}$

\begin{figure}[htbp]
  \begin{center}\parindent0mm
    \begin{tabular}{cccc}
      \epsfxsize=20mm\mbox{\epsfbox{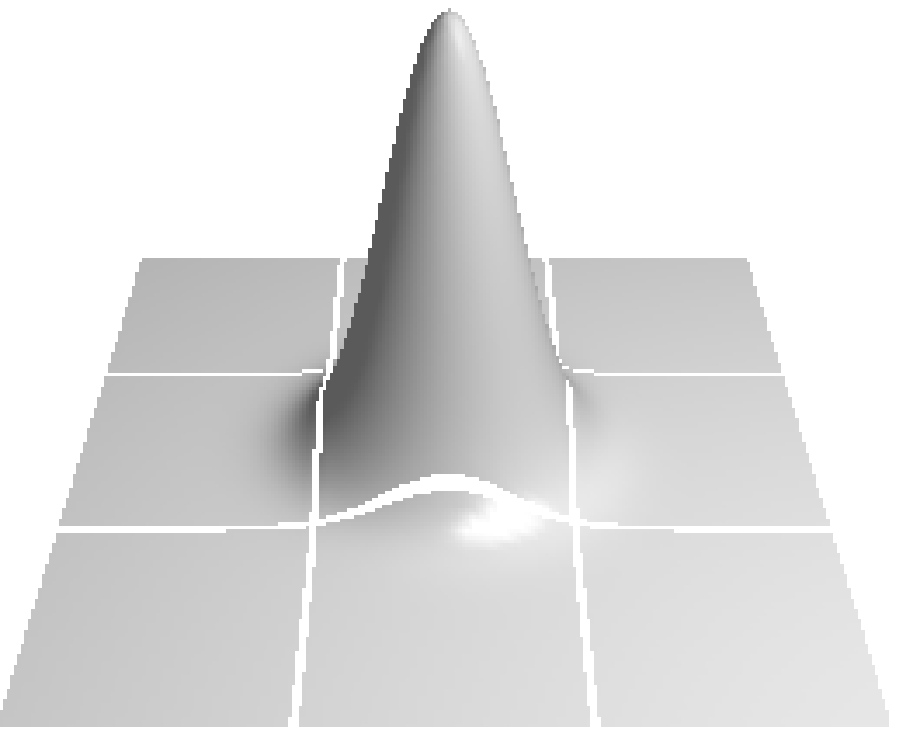}} &
      \epsfxsize=20mm\mbox{\epsfbox{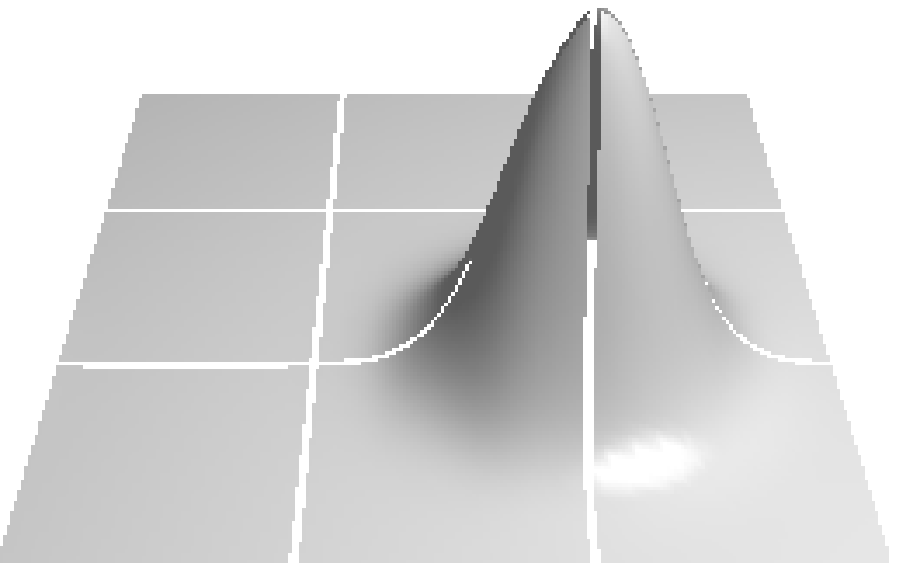}} &
      \epsfxsize=20mm\mbox{\epsfbox{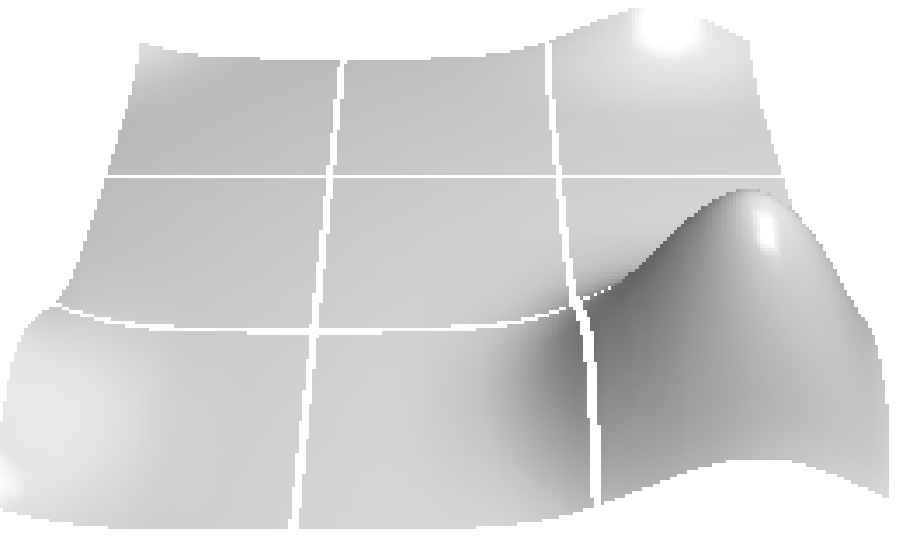}} &
      \epsfxsize=20mm\mbox{\epsfbox{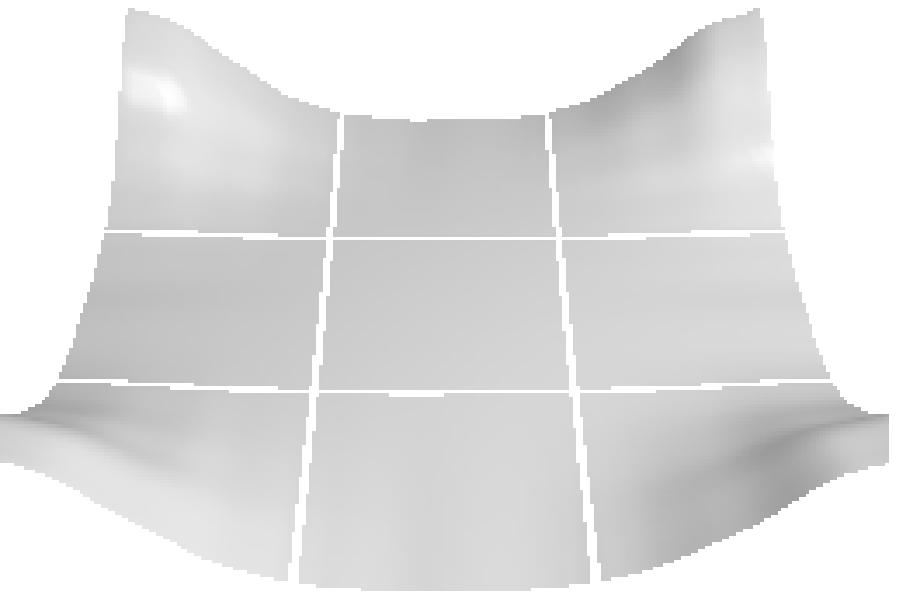}} \\
      \epsfxsize=20mm\mbox{\epsfbox{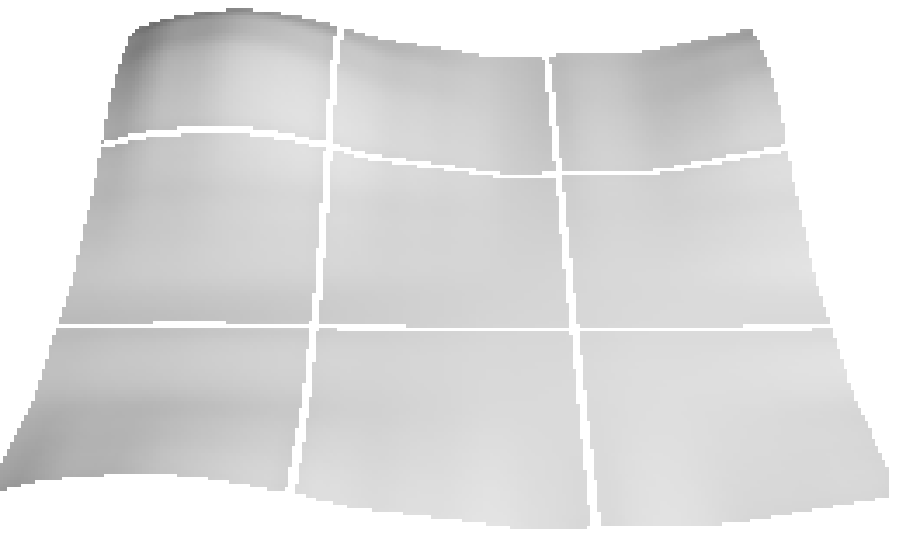}} &
      \epsfxsize=20mm\mbox{\epsfbox{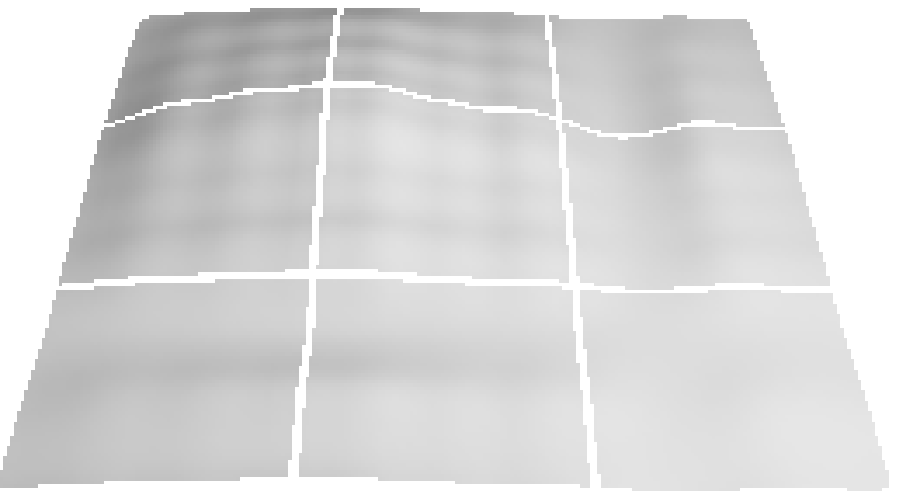}} &
      \epsfxsize=20mm\mbox{\epsfbox{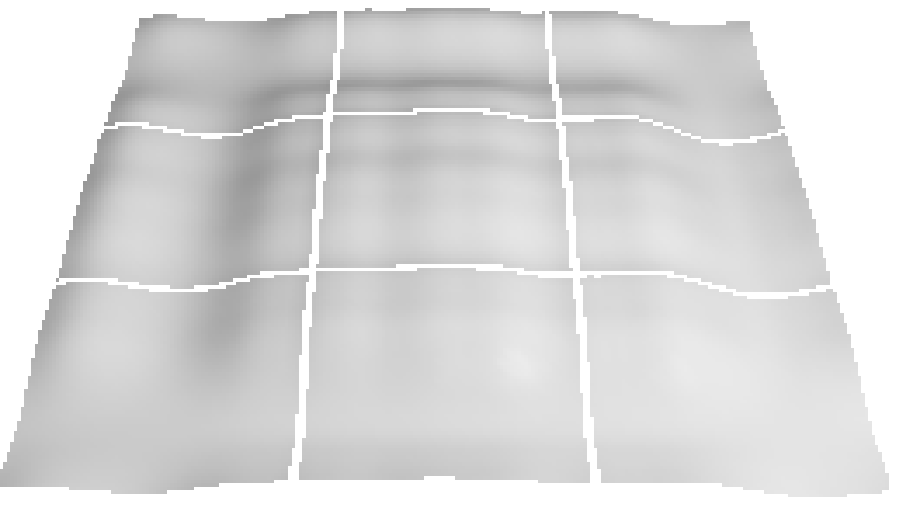}} &
      \epsfxsize=20mm\mbox{\epsfbox{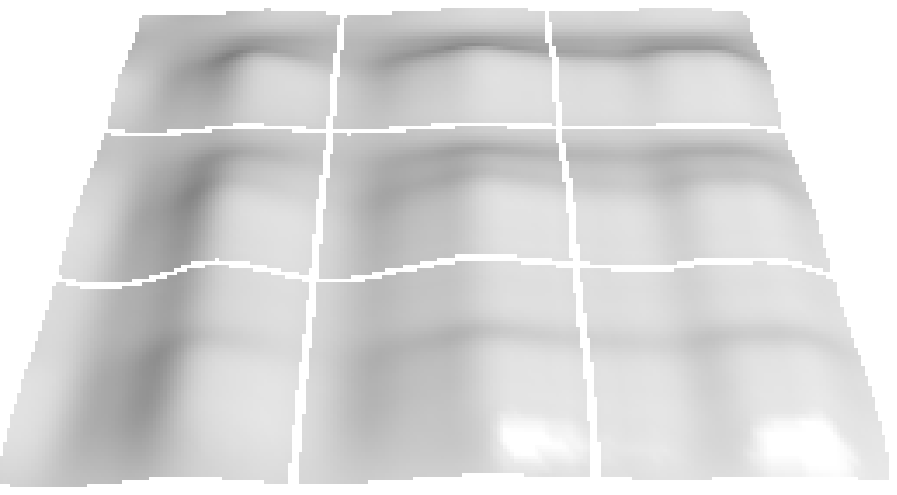}}
    \end{tabular}
  \end{center}
  \caption{
    Evolution of the density.
    The Gaussian is observed to go through these areas smoothly.
  }
  \label{fig3-2-2}
\end{figure}

\begin{figure}[htbp]
  \begin{center}
    \epsfxsize80mm\mbox{\epsfbox{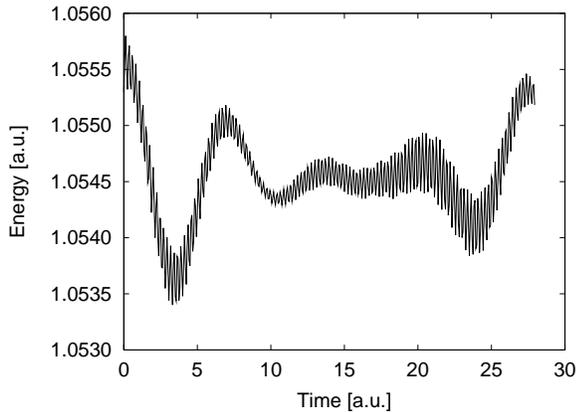}}
  \end{center}
  \caption{%
    Time variance in the energy.
    The initial energy is theoretically derived as $1.0625\text{a.u.}$,
    but it is evaluated as $1.0553\text{a.u.}$ by the FDM.
    The energy oscillates near its initial value
    but never drifts monotonously.
  }
  \label{fig3-2-3}
\end{figure}

Figure~\ref{fig3-2-2} shows snapshots of the time evolution of the Gaussian,
which is observed to go through these areas smoothly.
Figure~\ref{fig3-2-3} shows the evolution of the energy,
which is observed to oscillate around its initial value.

\begin{figure}[htbp]
  \begin{center}
    \epsfxsize=60mm\mbox{\epsfbox{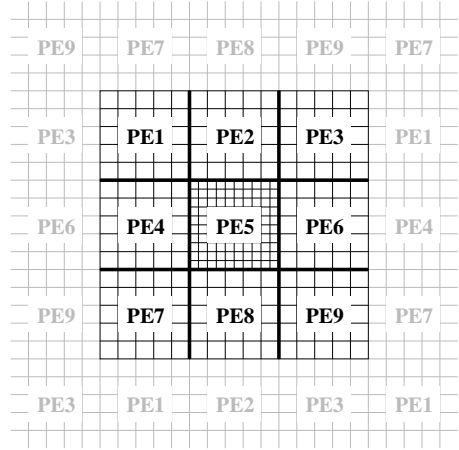}}%
  \end{center}
  \caption{
    The model system for the test of the adhesive operator
    for the adaptive mesh refinement.
    This system is also periodically connected and is divided into nine areas.
    Each area is managed by a single processing element.
    The size of each area is set at $8.0\text{a.u.}\times 8.0\text{a.u.}$\,
    $32\times 32$ computational grid points are allocated in each areas except
    the central area. The central area has $64\times 64$ computational
    grid points, which makes it twice as fine as those of the other areas.
    The time slice is set at $\Delta{t}=1/16\text{a.u.}$
  }
  \label{fig3-2-4}
\end{figure}

\begin{figure}[htbp]
  \begin{center}\parindent0mm
    \begin{tabular}{cccc}
      \epsfxsize=20mm\mbox{\epsfbox{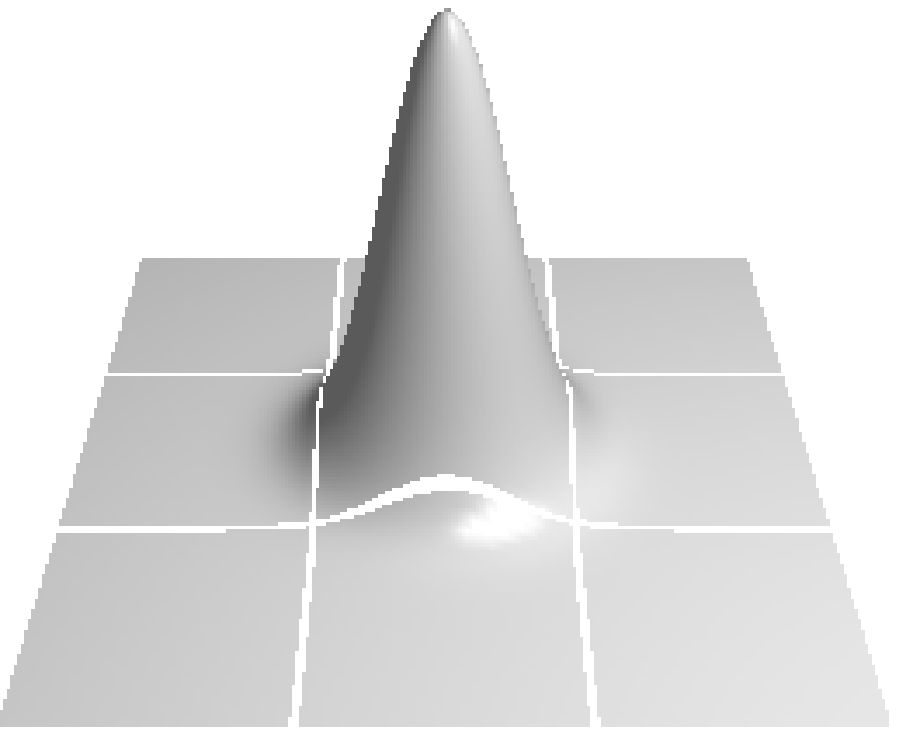}} &
      \epsfxsize=20mm\mbox{\epsfbox{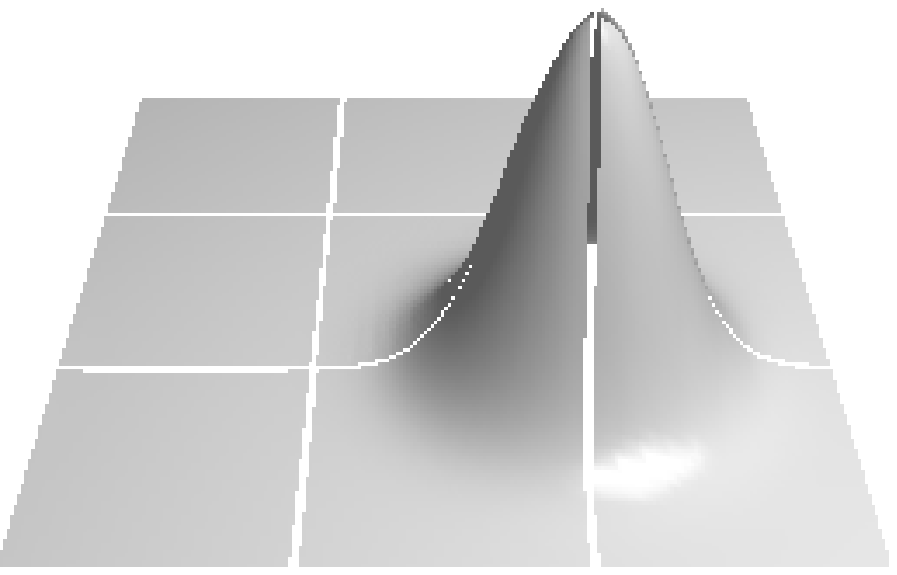}} &
      \epsfxsize=20mm\mbox{\epsfbox{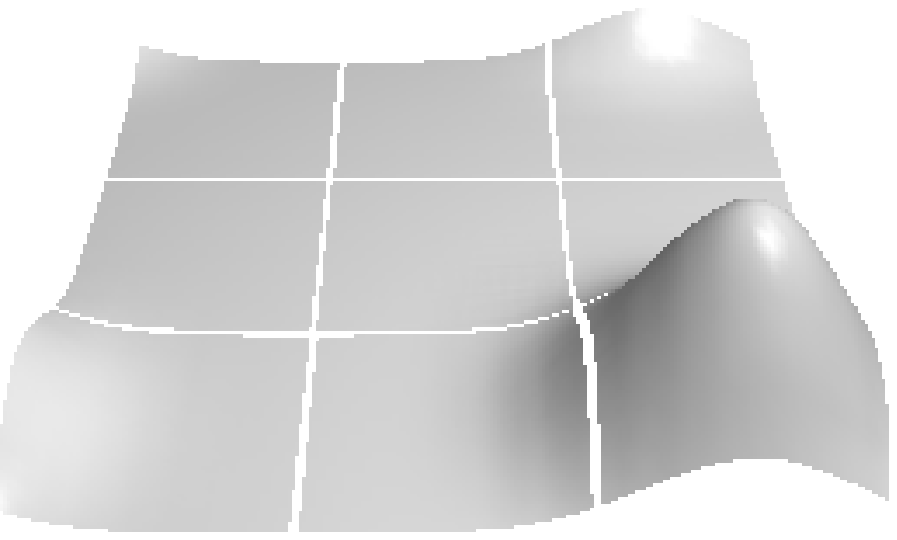}} &
      \epsfxsize=20mm\mbox{\epsfbox{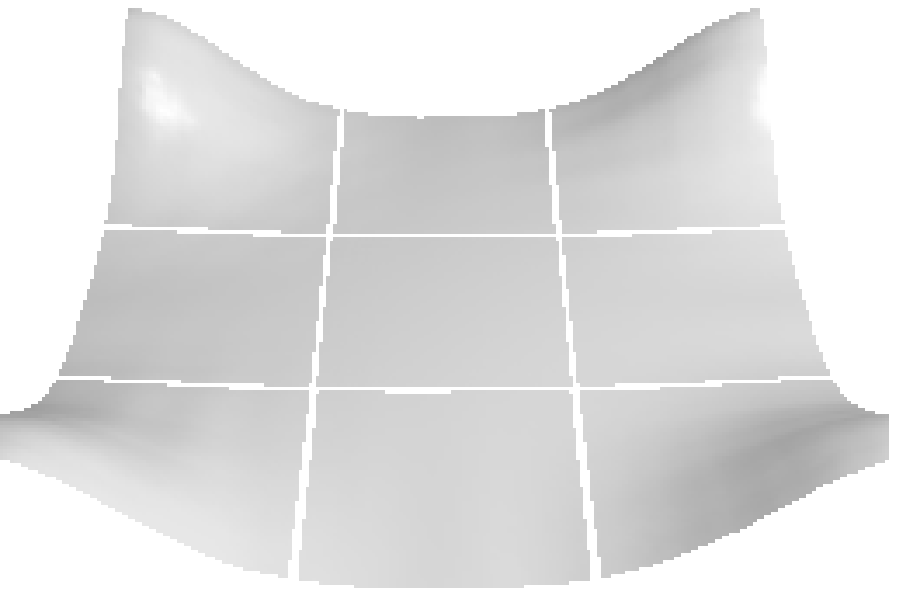}} \\
      \epsfxsize=20mm\mbox{\epsfbox{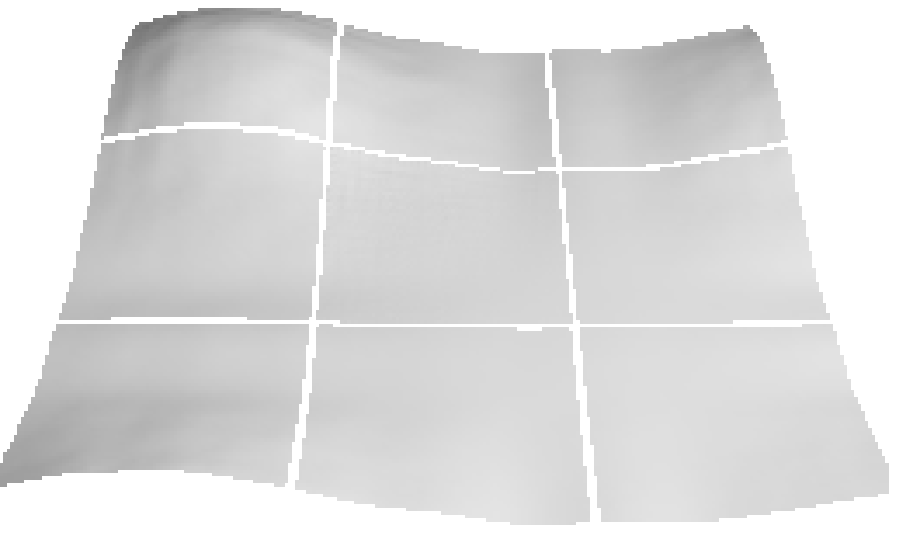}} &
      \epsfxsize=20mm\mbox{\epsfbox{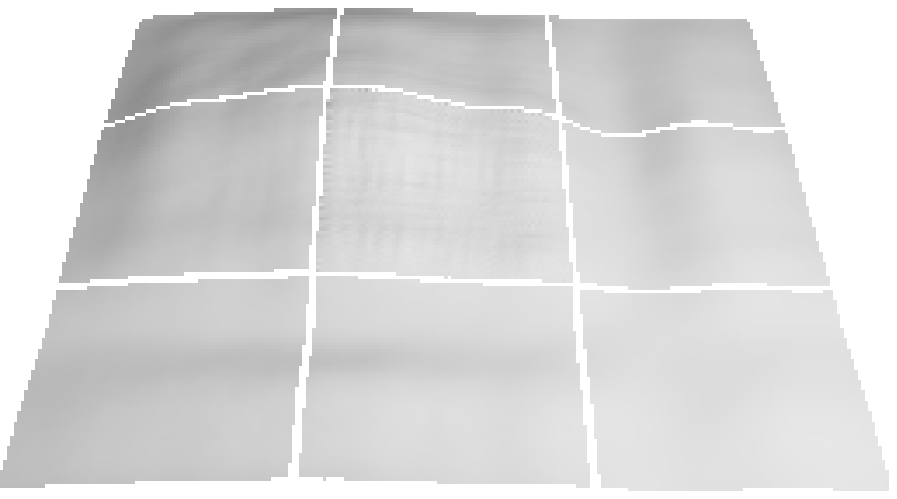}} &
      \epsfxsize=20mm\mbox{\epsfbox{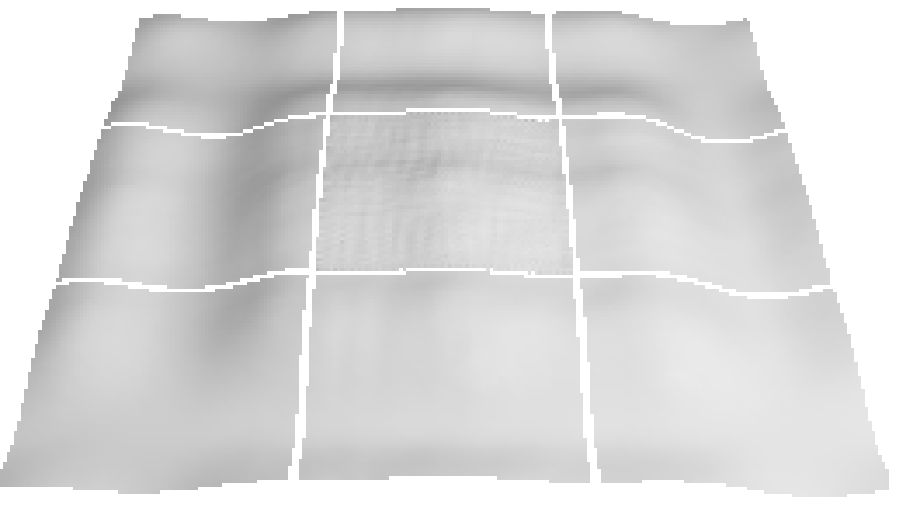}} &
      \epsfxsize=20mm\mbox{\epsfbox{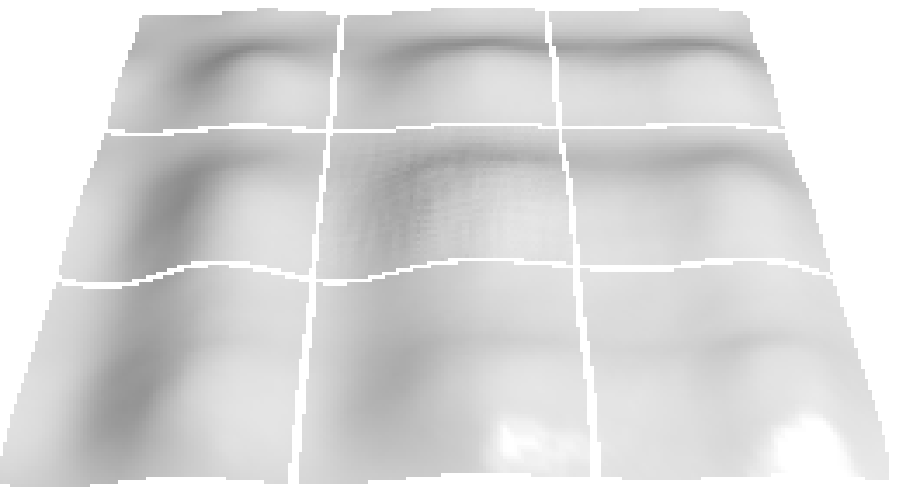}}
    \end{tabular}
  \end{center}
  \caption{
    Evolution of the density.
    The Gaussian is observed to go through these areas smoothly.
  }
  \label{fig3-2-5}
\end{figure}

\begin{figure}[htbp]
  \begin{center}
    \epsfxsize80mm\mbox{\epsfbox{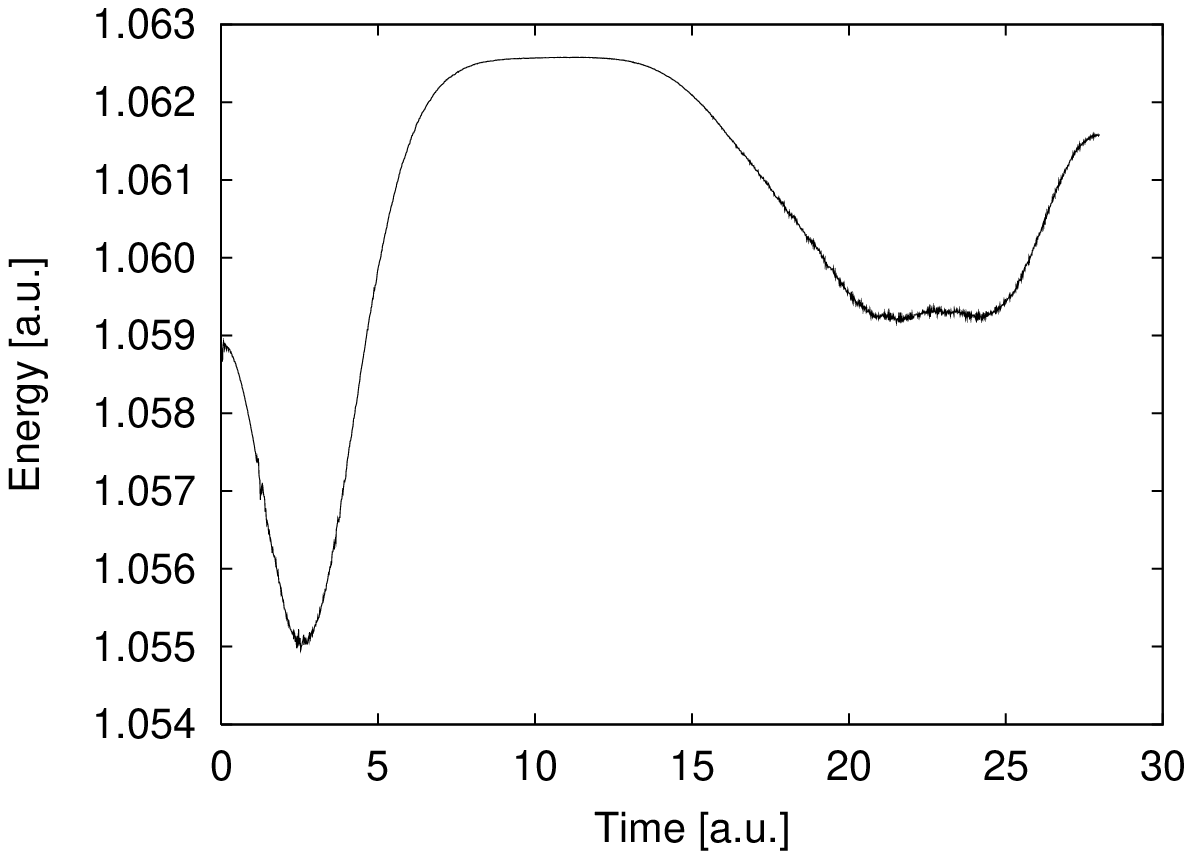}}
  \end{center}
  \caption{%
    Time variance in the energy.
    The initial energy is theoretically derived as $1.0625\text{a.u.}$,
    but it is evaluated as $1.0591\text{a.u.}$ by the FDM.
    The energy oscillates near its initial value
    but it never drifts monotonously.
  }
  \label{fig3-2-6}
\end{figure}

Second, we allocate $64\times 64$ grid points only in the central area
as illustrated in Fig.~\ref{fig3-2-4}.
We utilize the adhesive operator for the adaptive mesh refinement.
Figure~\ref{fig3-2-5} shows the snapshots,
with the Gaussian going through these areas smoothly.
Figure~\ref{fig3-2-6} shows the evolution of the energy,
which is observed to oscillate near its initial value.
In this way, the reliability of the adhesive operator is proved.

\subsection{Excitation of a hydrogen}

As the last application of the present method,
we demonstrate its validity and efficiency in describing
the process of photon-induced electron excitation
in a hydrogen atom in a strong laser field.
The laser is treated as a classically oscillating electric force
polarized in the z-direction:
\begin{equation}
  E_z = E_o \sin{\omega t}\ .
\label{3-7-1}
\end{equation}

The spatial variation of the electric field of the light is neglected, because
the electron system is much smaller than the order of the wave length.
Then the interaction term of the Hamiltonian is approximated as
\begin{equation}
  {\cal H}_{\rm int} = - e E_z z\ .
\label{3-7-2}
\end{equation}
In other words, we only take into account the electro-dipole
interaction of the electron with the light, and neglect the electro-quadrapole,
the magnetic-dipole, and other higher interactions.

The amplitude $E_o$ is set at $1/64\text{a.u.}=0.80\text{V/\AA}$,
which is as strong as
a usual pulse laser. The angular frequency $\omega$ is set at
$0.3125\text{a.u.}=8.5\text{eV}$,
less than the transition energy between 1S and 2P. Ordinarily,
such low energetic electric force has no effect on the electronic excitation.
But with such a strong amplitude, various nonlinear optical effects
are caused by the electron dynamics.

We allocate $128^3$ grid points in a $32^3\text{a.u.}^3$ cubic closed system.
The hydrogen nucleus is located at the center of the system,
and the nucleus potential is constructed by solving the Poisson equation
in the discretized space to avoid the singularity of the nucleus potential.
The 1S-orbital is assumed as the initial state of the wavefunction.
Then we turn on the electric field and start the simulation.
The time slice is set at $0.0785\text{a.u.}=2.0\times 10^{-3}\text{fs}$ so as to
follow the rapid variation of the wavefunction and the electric force.
We follow the evolution for $32\text{k}$ iteration.

\begin{figure}[htbp]
  \begin{center}
    \epsfxsize80mm\mbox{\epsfbox{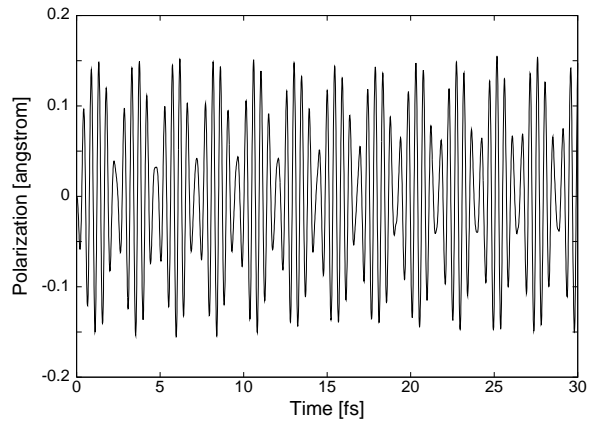}}
  \end{center}
  \caption{
    Time variance in the polarization of the electron.
  }
  \label{fig3-7-3}
\end{figure}

\begin{figure}[htbp]
  \begin{center}
    \epsfxsize80mm\mbox{\epsfbox{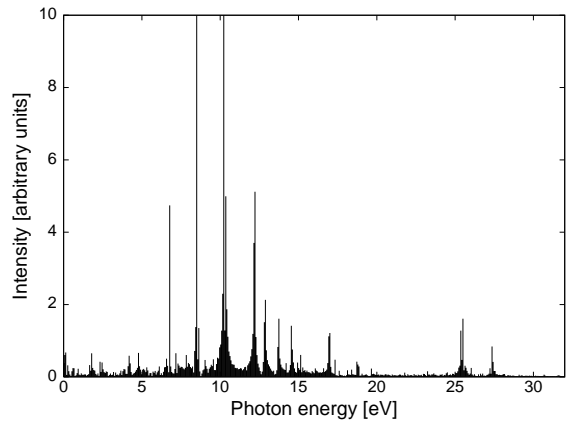}}
  \end{center}
  \caption{
    Spectrum of the scattered light
    generated by the oscillation of the electron.
  }
  \label{fig3-7-4}
\end{figure}

Figure~\ref{fig3-7-3} shows the time variance in the polarization of the electron.
The oscillation of the polarization generates another electric field,
which corresponds to a non-linearly scattered light from the atom.
By Fourier-transforming the polarization along the time axis, we obtained
the spectrum of the scattered light shown in Fig.~\ref{fig3-7-4}.

Several sharp peaks are found, which are interpreted as follows:
The peak at $8.5\text{eV}$ comes from Rayleigh scattering,
whose frequency is identical with the injected light: $\omega$.
The peak at $10.2\text{eV}$ comes from Lyman $\alpha$ emission,
which is generated by the electron transition from
the 2P-orbital to the 1S-orbital: $\omega_{L_\alpha}$.
On the other hand, the peak at $12.1\text{eV}$ comes from Lyman $\beta$
emission, which is generated by the electron transition from
the 3P-orbital to the 1S-orbital: $\omega_{L_\beta}$.
The peak at $6.8\text{eV}$ comes from hyper Raman scattering,
whose frequency is identical with $2\omega-\omega_{L_\alpha}$.
Moreover the peak at $25.5\text{eV}$ comes from the third harmonic
generation, whose frequency is identical with $3\omega$.

\begin{figure}[htbp]
  \begin{center}\parindent0mm
    \begin{tabular}{cccc}
      \epsfxsize=20mm\mbox{\epsfbox{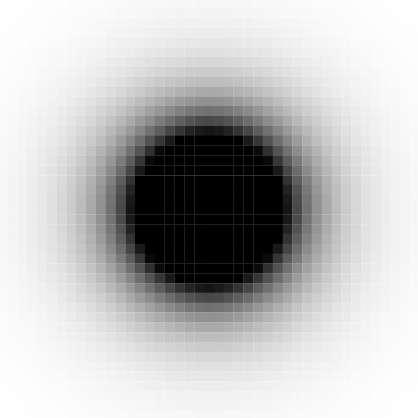}} &
      \epsfxsize=20mm\mbox{\epsfbox{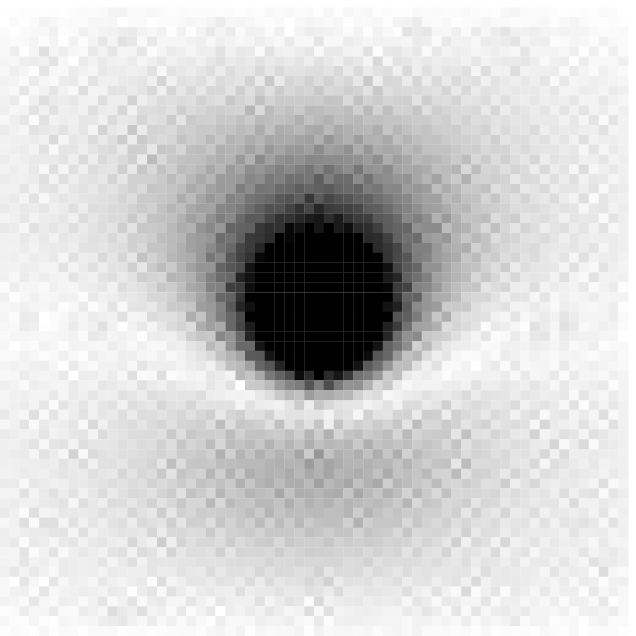}} &
      \epsfxsize=20mm\mbox{\epsfbox{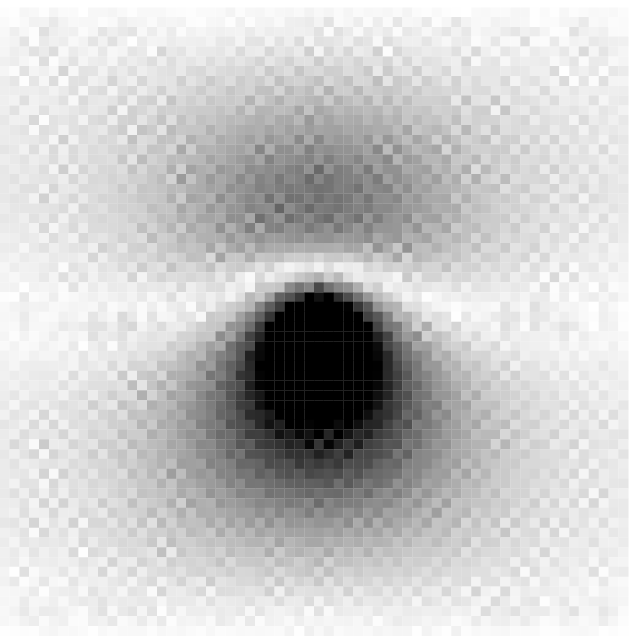}} &
      \epsfxsize=20mm\mbox{\epsfbox{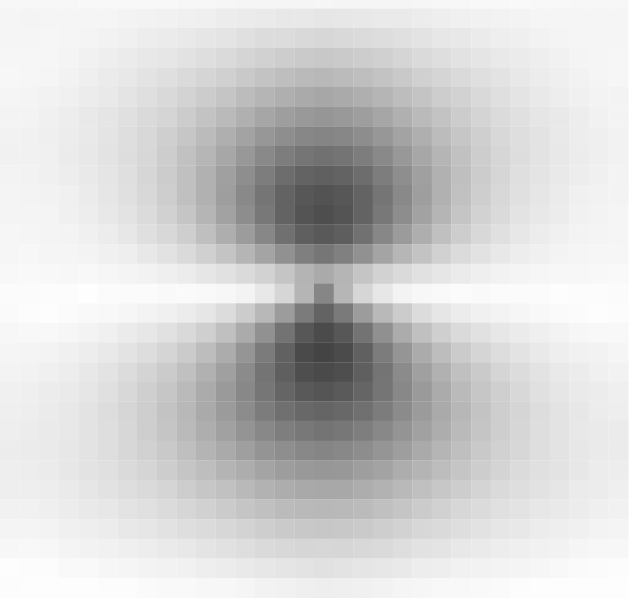}}
    \end{tabular}
  \end{center}
  \caption{
    Evolution of the density of the electron in the hydrogen atom.
    The density starting from a 1S orbital 
    oscillates with time and becomes a 2Pz orbital.
  }
  \label{fig3-7-2}
\end{figure}

The simulation is also performed for a different laser frequency;
the injecting photon energy $\omega$ is set at $10.2\text{eV}$,
which is the same as the transition energy between 1S and 2P.
In this case the electron starting from a 1S orbital 
is expected to excite to a 2Pz orbital. Figure~\ref{fig3-7-2}
shows the snapshots of the density during the simulation time span.

\begin{figure}[htbp]
  \begin{center}
    \epsfxsize70mm\mbox{\epsfbox{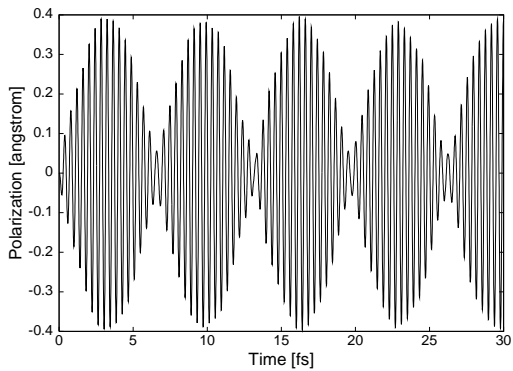}}
  \end{center}
  \caption{
    Time variance in the polarization of the electron.
  }
  \label{fig3-7-5}
\end{figure}

\begin{figure}[htbp]
  \begin{center}
    \epsfxsize70mm\mbox{\epsfbox{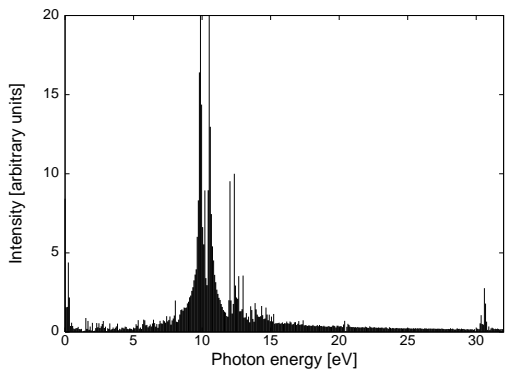}}
  \end{center}
  \caption{
    Spectrum of the scattered light
    generated by the oscillation of the electron.
  }
  \label{fig3-7-6}
\end{figure}

Figure~\ref{fig3-7-5} and Fig.~\ref{fig3-7-6}
show the polarization and the spectrum, respectively.
Three peaks are found, at $9.9\text{eV}$, $10.2\text{eV}$, and $10.5\text{eV}$.
These peaks are derived from the theory of the Dressed atom or the 
AC stark effect as below:
\begin{equation}
 \omega - eE_o \left<{\rm 2P_z}|z|{\rm 1S}\right>,\
 \omega,\ 
 \omega + eE_o \left<{\rm 2P_z}|z|{\rm 1S}\right>\ .
\label{3-7-3}
\end{equation}

One could obtain such behavior analytically by
using perturbation theory; however, with the present method,
we could directly calculate them without perturbation theory
and without information on the excited states of the system.

\section{Conclusion}

We have formulated a new method for solving the time-dependent
Schr{\"o}dinger equation numerically in real space.
We have found that by using Cayley's form and Suzuki's
fractal decomposition, the simulation can be fast, stable, accurate,
and suitable for vector-type supercomputers. We have proposed the adhesive
operator to make Cayley's form suitable for periodic systems and
parallelization and adaptive mesh refinement.

These techniques will also be useful for the time-dependent Kohn Sham equation,
which is our future work.

\section{Acknowledgments}

We are indebted to Takahiro Kuga for his suggestions concerning
non-linear optics. Calculations were done using the SR8000
supercomputer system at the Computer Centre, University of Tokyo.


\end{multicols}

\end{document}